\documentclass[a4paper]{article}

\usepackage{graphicx}
\usepackage[english]{babel}
\usepackage{amssymb,amsthm,amsmath,latexsym,amsfonts,epsfig}
\usepackage{subfigure} 
\usepackage{multirow}

\usepackage{float}
\usepackage[section]{placeins}
\usepackage{pstricks}

\def\beq{\begin{equation}}
\def\eeq{\end{equation}}
\def\bea{\begin{eqnarray}}
\def\bean{\begin{eqnarray}}
\def\eea{\end{eqnarray}}
\def\R{{\mathbb R}}

\def\N{{\mathbb N}}

\newcommand{\ind}{\indent}

\newcommand{\non}{\nonumber}
\newcommand{\dps}{\displaystyle}

\newcommand{\dif}{\mathrm{d}}
\newcommand{\im}{\mathrm{i}}
\newcommand{\sign}{\mathrm{sign}}
\newcommand{\p}{\mathrm{p}}

\definecolor{verdeoscuro}{rgb}{0,0.7,0}

\begin{document}

\begin{center}
{\LARGE{\bf{Quantum mechanics and umbral calculus}}}
\end{center}

\bigskip\bigskip

\begin{center}
J E L\'opez-Sendino$^1$, J Negro$^2$, M A del Olmo$^3$
and E Salgado
\end{center}

\begin{center}
Departamento de F\'{\i}sica Te\'orica, Universidad de
Valladolid,\\
E-47011, Valladolid, Spain
\medskip

{e-mail: $^1$jels@metodos.fam.cie.uva.es, $^2$jnegro@fta.uva.es, $^3$olmo@fta.uva.es}
\end{center}
\bigskip

\begin{abstract}
In this paper we present the first steps for obtaining a discrete Quantum Mechanics making use of
the Umbral Calculus. The idea is to discretize the continuous Schr\"odinger equation substituting
the
continuous derivatives by discrete ones and the space-time continuous variables by well
determined operators that verify some Umbral Calculus conditions. In this way we assure that 
some properties of integrability and symmetries of the continuous equation are preserved and also
the solutions of the continuous case can be recovered discretized in a simple way.
The case of the Schr\"odinger equation with a potential depending only in the space variable is
discussed.
\end{abstract}


\section{Introduction}

In the last years the interest in physics for discrete space-time theories has increased mainly  in
relation with quantum gravity \cite{gibbs}. In a scenario where the physical space-time may be
discrete (i.e. there is a fundamental length in the space-time \textendash  let
$\sigma$ be the fundamental space length and $\tau$ the fundamental time length \textendash\, that
could be related with Planck's constant), continuous theories would only be approximations to the
reality. Hence, it seems interesting the study of discrete physical theories such that in the limit
of $\sigma$ going to zero we recover the well known and established continuous physical theories.
One way to do that is to discretize continuous physical theories, in particular  quantum theories.
It is obvious, that after discretization some properties will be conserved  and other ones will
disappear. A detailed  study about when, how and why this happens would be pertinent and very
interesting. This is the aim of this work: to study how to discretize a physical theory in such a
way that properties related with symmetries in the continuous case can be preserved in the process,
and so, to analyse the behaviour of quantum mechanics after discretization.

The Umbral Calculus is the mathematical tool that we will use to discretize Quantum Mechanics. It
has been used recently to provide discrete representations of canonical com\-mu\-ta\-tion relations 
(i.e. $[\partial_x,x]=1$) in order to discretize linear differential equations
\mbox{\cite{dimakis}--\cite{salgado}}, in such a way that the continuous point symmetries are
preserved the most cases. 

In this paper we will consider a fixed lattice and use the umbral correspondence to obtain discrete
solutions maintaining certain commutation relations \cite{salgado}. 

The history of the Umbral Calculus goes back to the 17th century: Taylor, Newton, Barrow,
Vandermonde, etc.  \cite{mullin} found some theorems and expressions relating
powers with polynomials. However, in the second half of the 19th century through Sylvester, Cayley
and Blissard \mbox{\cite{blissard1}--\cite{bell}} appeared the terms `umbrae' (shadow in Latin) and
`Umbral Calculus' for the first time in relation with a set of `magic' rules of lowering and raising
indices. Some mathematical developments, like the theory of Sheffer polynomials (called
Appell polynomials or more generally polynomials of type zero) \mbox{\cite{appell}--\cite{sheffer}}
and the theory of abstract linear operators by Pincherle \cite{pincherle}, converged to Umbral
Calculus. Finally, in the second half of the 20th century  Rota, Roman and collaborators
\mbox{\cite{rota1}--\cite{roman5}} developed the Umbral Calculus as a linear algebra of
operators. Thus, they formulated the theory of the umbral algebra and its dual in the space of
formal series. A large bibliography about Umbral Calculus and its applications can be found in
Ref.~\cite{bucchianico}.

About recent results, we can mention that new methods in Umbral Calculus have been introduced for
computation of series by integration \cite{zachos}. Also
Ref.~\mbox{\cite{levi3}--\cite{levi7}} study the symmetries of nonlinear difference equations or
multidimensional ones related with Quantum Mechanics.

This paper is organised as follows. In the next section we introduce the basic concepts of
Umbral Calculus in terms of the theory of linear operators. In section~\ref{umbralcorrespondence}
we show the main facts relative to the discretization of the (continuous) Quantum Mechanics. The
most usual representations for the discrete derivative operator (right, left ¡and symmetric discrete
derivatives) are analysed in section~\ref{differentialoperators}. In order to see how the above
umbral correspondences work we present in section~\ref{discreteexponential} their action on some
non-polynomial functions: the exponential and the trigonometric functions. The next section
is devoted to present the basis for constructing a discrete Quantum Mechanics founded on the Umbral
Calculus. Thus we present an umbral version of the Schr\"odinger equation for some
potentials. Finally some conclusions close the paper.


\section{Introduction to Umbral Calculus}\label{umbralcalculus}

This introductory exposition of the modern theory of the Umbral Calculus, developed mainly by Rota 
and Roman, follows mainly  Ref.~\cite{roman5}. The {\it umbral algebra}, $\mathcal{L}$, is the
algebra of linear operators acting on the algebra, $\mathcal{F}$, of formal power series in a
variable or acting on  the algebra of real or complex polynomials in a variable, $\mathcal{P}$. 
The elements of $\mathcal{F}$ are of the form
\[
f(x)=\sum_{n=0}^\infty a_n x^n, \quad  a_n\in \mathbb{F} ,
\]
and those of $\mathcal{P}$ are
\[
p_n(x)=\sum_{i=0}^n a_i x^i, \quad  a_i\in \mathbb{F},\; n\in \mathbb{N},
\]
being $\mathbb{F}$ a field of characteristic zero (in our case it will be identified with
the real numbers $\mathbb{R}$, but it could be, for example, the complex numbers).

The linear operators of $\mathcal{L}$ called {\it shift operator}, $T$, and {\it coordinate
operator}, $X$,   act on the elements of $\mathcal{F}$ or $\mathcal{P}$ as follows
\beq
\begin{array}{l}
T_\sigma\cdot p(x)= p(x+\sigma),\\[0.3cm]
X\cdot p(x)=x\: p(x).
\end{array}
\label{eqX}
\eeq
All the results of this work can be  extended to separable equations of any number of
variables. For non separable equations the method is the same but the results may be quite
different.

An operator $O\in\mathcal{L}$ is said {\it shift invariant} if
and only if commutes with the shift operator, i.e., 
\[
[O,T_\sigma]=O T_\sigma-T_\sigma O=0 , \qquad 
\forall\:\sigma\in\mathbb{F} .
\]

The so-called {\it delta operators}, $\Delta$,  are those operators  of 
$\mathcal{L}$ such as
\beq
\begin{array}{l}
\Delta\cdot x= c\neq0 ,\qquad  c\in\mathbb{F}, \\[0.30cm]
[\Delta,T_\sigma]=\Delta T_\sigma-T_\sigma \Delta =0 ,\qquad
\forall\:\sigma\in\mathbb{F},
\end{array}
\label{eqdeltaop}
\eeq
where $c\in\mathbb{F}$ is any fixed constant. It is easy to prove that
\[
\Delta\cdot a =0, \qquad \forall\
a\in\mathbb{F}.
\]
For every  delta operator $\Delta$  there is a {\it series of basic polynomials} \newline$\{
p_0(x), p_1(x),p_2(x), \dots, p_n(x) \}$ of $\mathcal{P}$ verifying
\beq
\begin{array}{l}
p_0(x)=1 ,\\[0.3cm]
p_n(0)=0  , \quad \forall\ n> 0, \\[0.3cm]
\Delta\cdot p_n(x)=n\ p_{n-1}(x) .
\end{array}
\label{eqcondmala}
\eeq
We see that  $\Delta$  acts as a lowering (index) operator on its associated polynomial. 
A very important fact of the theory is that each pair $\{\Delta,p_n(x)\}$
defines a unique realization or representation of the umbral algebra. There is a bijective correspondence
$\mathfrak{J}$, between basic series of polynomials and their $\Delta$   operators. Hence,
$\mathfrak{J}$ may be defined with only one element of the pair.

The commutator of an operator $O$  with the coordinate operator $X$
\beq
O'=[O,X]=O X-X O
\label{eqpinchderiv}
\eeq
is called  the {\it Pincherle derivative} of $O$.

Given a delta operator $\Delta$ we can find an associated operator 
$\xi$ such that
\beq
[\Delta,\xi]=1.
\label{heisenberg}
\eeq
This operator $\xi$ is not shift invariant and together with $\Delta$ and $1$ close the
Heisenberg-Weyl algebra or, in other words, expression (\ref{heisenberg}) is a Heisenberg-like
relation between 
$\Delta$ and its conjugate $\xi$.
It is worthy to note that $\xi$ is associated with the corresponding 
basic series of the operator $\Delta$  by
\[
\xi^n\cdot 1=p_n(x).
\]
Although only in a determined correspondence $\xi^n\cdot 1$ coincides with the Pockhammer symbol
$x^{(n)}$, from now on we will use the following notation 
\beq
\xi^n\cdot 1=p_n(x)\equiv x^{(n)} ,
\label{basicseries}
\eeq
which is easy to prove  taking into account that 
\[
[\Delta,\xi^n]=n\;\xi^{n-1} .
\]
Obviously, there is only one way to define $\xi$ if its associated polynomial series obeys all the
relations (\ref{eqcondmala}). Given an operator $\Delta$ its associated operator $\xi$  must be
\[
\xi=X\beta ,
\]
where $\beta^{-1}$ is the first Pincherle derivative of $\Delta$. It is easy to demonstrate that
$\beta^{-1}$ is invertible and $\beta$ is shift invariant.

Let  $\Delta_1$ and $\Delta_2$ be  two delta operators with associated  $\xi$-operators $\xi_1$ and
$\xi_2$, respectively. A map $\mathcal{R}:\mathcal{L}\rightarrow\mathcal{L}$ is called an {\it
umbral correspondence} if
\[
\mathcal{R}: \xi_1^n\;\rightarrow\;\xi_2^n, \qquad \forall n\in \N.
\]
The umbral correspondence also induces a correspondence $\mathcal{R}$ between two  delta operators.
Summarising, 
the following  diagram is commutative
\[
\begin{array}{c@{}c@{}c}
\:\:\xi_1^n             &\overset{\mathcal{R}}{\longleftrightarrow}&\:\:\xi_2^n\\[3pt]
\mathfrak{J}\updownarrow&                                          &\mathfrak{J}\updownarrow\\[3pt]
\:\:\Delta_1            &\overset{\mathcal{R}}{\longleftrightarrow}&\:\:\Delta_2\\[3pt]
\mathfrak{J}\updownarrow&                                          &\mathfrak{J}\updownarrow\\[3pt]
\xi_1^n\cdot1=x_1^{(n)} &\overset{\mathcal{R}}{\leftrightarrow}    &\xi_2^n\cdot1=x_2^{(n)}
\end{array}
\]
From the above diagram one easily sees that there is a triplet $\{\Delta,\xi,p_n(x)\}$ that defines
a unique representation of the umbral algebra. Moreover through $\mathfrak{J}$ we can go in a
bijective way from one element to another of the triplet in such a way that only one of these
elements is necessary and sufficient to define the umbral representation. The bijective application
$\mathcal{R}$ bring us from a umbral representation to another one.


\section{Umbral correspondence and discretized quantum mechanics}
\label{umbralcorrespondence}

As we mentioned before, the main goal of this work is the application of the umbral correspondence
to discretize
Quantum Mechanics. In fact, the 1-dimensional Quantum Mechanics is an umbral realization via the
identification
\[
\Delta=\partial_x  , \qquad \xi=X,  \qquad p_n(x)=x^n.
\]
Since we are interested in the discretization of Quantum Mechanics we will use the umbral
correspondence to relate the triplet $(\partial_x , \; X, \; x^n)$ with another one defined
on the lattice field $\mathbb{F}_{\sigma}=\{x\in\mathbb{R}\ | x=m \sigma, \; m\in\mathbb{Z}\}$
with lattice parameter $\sigma\in\mathbb{R}$.

Let us start defining a {\it general difference operator}, $\Delta$, as a delta operator such that
in the limit when $\sigma$ goes to zero becomes the differential operator $\partial_x$. Supposing
that $\Delta$ is a linear difference operator then it will have a polynomial dependence on the
shift operator $T$, i.e.,
\beq
\Delta=\frac{1}{N\sigma}\sum_{n=-j}^k a_nT^n ,
\label{generaldifference}
\eeq
where $N \in \N$, $a_n\in \R$ and $\sigma\in\mathbb{R}$ is the lattice
parameter.
Imposing condition 
(\ref{eqdeltaop}a), i.e. $\Delta\cdot x=1$,  since  condition (\ref{eqdeltaop}b) is trivially verified we get
\beq
 \begin{array}{l}
\sum_{n=-j}^k a_n=0, \\[0.3cm]
\sum_{n=-j}^k n a_n=N.
\end{array} 
\label{eqconddiffe}
\eeq
These conditions (\ref{eqconddiffe}) guarantee that in the limit when $\sigma$ goes to zero $\Delta$
goes to $\partial_x$ as it is easy to prove.

Recalling that the pair of Quantum Mechanics operators
$\left(\partial_x, X\right)$ verifies the Heisenberg relation 
\[
\left[\partial_x, X\right]=1,
\]
we call {\it discrete position operator} to a non-shift invariant operator $\xi$ such that with
the difference operator $\Delta$ verifies  a similar Heisenberg relation, i.e.,   
\[
[\Delta,\xi]=1.
\]

The umbral correspondence allows us to relate continuous linear equations $h$ and their solutions
$f$ if they can be developed in power series, with umbral discrete versions $\hat{h}$ and $\hat{f}$,
respectively, as follows
\beq
\begin{array}{lll}
\dps h\left(\partial^n_x f(x),x^s\right)=0&\longleftrightarrow &
\dps\hat{h}\left(\Delta^n \hat{f}(\xi),\xi^s\right)\cdot1=0 ,\qquad \: n,s\in\mathbb{N};\\[0.4cm]
\dps f(x)=\sum_{n=0}^\infty f_n x^n &\longleftrightarrow &
\dps \hat{f}(\xi)\cdot1=\sum_{n=0}^\infty f_n x^{(n)} .
\end{array}
\label{generalcorrespondence}
\eeq
This is the main result of Umbral Calculus that we will exploit here for ours purposes. Umbral
Calculus allows us not only to discretize continuous differential equations obtaining discrete
equations but also to get solutions of these equations if their continuous counterparts are
polynomial solutions or may be developed in power series which converge after the correspondence.
Note, however, that with the umbral correspondence we obtain an operator which must be applied to
$1$ to get a difference equation or its solution.

For the particular case of a {\it symmetric difference operator}, i.e., a difference
operator $\Delta_\sigma$ such that 
\[
\Delta_\sigma \cdot f(x)=-\Delta_{-\sigma} \cdot f(x),
\]
we have that the general expression (\ref{generaldifference}) reduces to 
\beq
\Delta_\sigma=\frac 1{N\sigma}\sum_{n=1}^j a_n (T^n-T^{-n}).
\label{nsymmetric}
\eeq


\section{Differential operators for discretized quantum mechanics}
\label{differentialoperators}

In the following we will use some particular expressions (or representations) for the differential
operator $\Delta$, all of them of low (first and second) order on the shift operator $T$. Let us
consider the two possible representations for $\Delta$ of first order,  the right discrete
derivative $\Delta_+$ and the left discrete derivative $\Delta_-$, and the (only) symmetric discrete
derivative  of second order  $\Delta_s$. 

The {\it right-correspondence} is determined by the couple of operators $(\Delta_+,\xi_+)$ given by
\[
\Delta_+:=\frac{1}{\sigma}(T-1), \qquad
\beta_+ =T^{-1},
\]
and the {\it left-correspondence} $(\Delta_-,\xi_-)$ by
\[
\Delta_-:=\frac{1}{\sigma}(1-T^{-1}), \qquad
\beta_- =T .
\]
Note that both are connected by the change $\sigma \leftrightarrow -\sigma$.

The basic polynomial series of the right-correspondence is related with the Pochhammer symbols. Its
explicit expression is 
\beq
\begin{array}{lll}
(X\beta_+)^n\cdot 1=x_+^{(n)}&=&
x(x-\sigma)(x-2\sigma)\dots(x-(n-1)\sigma)\\[0.3cm]
          &=& \prod_{i=0}^{n-1}(x- i\sigma )=
                  \sigma^n \prod_{i=0}^{n-1}(m-i)\\[0.4cm]
                    &=&\left\{\begin{array}{ll}
                     (-\sigma)^n \displaystyle\frac{(-m+n-1)!}{(-m-1)!}&\text{if } 
                     m<0\\
                     0&\text{if }\left\{\begin{array}{l}
                                  m\geq 0\\
                                  m<n
                                  \end{array}\right.\\
                     \sigma^n\displaystyle\frac{m!}{(m-n)!}&\text{if }\left\{\begin{array}{l}
                                  m\geq 0\\
                                  m\geq n
                                  \end{array}\right.
                  \end{array}\right.
\end{array}
\label{xbetaMas}
\eeq
where $x=m\sigma$ with $ m\in\mathbb{Z}$.
This series of polynomials has some interesting properties. First of all we can see that the $n$-th
order zero of $x^n$  splits into $n$ first order zeros of $x^{(n)}_+$ at the right of the origin.
If the right-correspondence is applied to a Taylor power series then, for the positive points of the
domain, it will be truncated and therefore will converge, but not for the negative ones, and
therefore could diverge.

The left-correspondence $\{\Delta_-,\xi_-\}$ is nearly similar to the right one. The basic
sequence is related with the rising factorial (instead of the falling factorial of the previous
case), and it is mirror symmetric respect to the y-axis to the right-correspondence: in fact, the
negative points of a Taylor series discretized in this way, always converge.

According to the formula (\ref{nsymmetric}) there is only one symmetric representation of second
order on $T$  ($S$-representation) determined by the pair $\{\Delta_s,\xi_s\}$, where
\[
\dps\Delta_s=\frac{1}{2\sigma}(T-T^{-1}),\qquad
\beta_s =2(T+T^{-1})^{-1},
\]
and its basic sequence is
\bea
(X\beta_s)^n\cdot1&=&x_s^{(n)}
=x[x-(n-2)\sigma][x-(n-4)\sigma]\dots[x+(n-4)\sigma][x+(n-2)\sigma]\non\\[0.3cm]
                  &=&x\prod_{i=0}^{n-2}[x+(2i-(n-2))\sigma]=\sigma^n
m\prod_{i=0}^{n-2}[m+2i-(n-2)]\non\\[0.3cm]
                  &=&\left\{\begin{array}{ll}
                     (\sign(m)\sigma)^n |m|\dps\frac{(|m|+n-2)!!}{(|m|-n)!!}&\text{if }n \leq |m|\\
                     0&\text{if }\left\{\begin{array}{l}
                                  n > |m|\\
                                  \p(n-m)=+1
                                  \end{array}\right.\\
                     \begin{array}{l}
(-1)^{\frac{1}{2}(n-|m|-1)}(\sign(m)\sigma)^n m\\
\quad\times (|m|+n-2)!!(n-|m|-2)!!\end{array}&\text{if }
\left\{\begin{array}{l}
                                  n> |m|\\
                                  \p(n-m)=-1
                                  \end{array}\right.
                  \end{array}\right.
\label{xbetasn}
\eea
where  $\p(n)=(-1)^n$ is the parity function and the double factorial is defined as $n!!=n (n-2)!!$
with $n!!=1$ when $n$ is $0$ or $1$.

The behaviour of this basic sequence is much more complicated than those of the right and left-correspondences.
The zero at the origin of order $n$ of the continuous power function of degree $n$ splits into $n$
simple zeros after discretization. These zeros appear at the points $x=m\sigma$ with the conditions 
\beq\label{condiciones}
\p(n-m)=+1,\qquad n>|m|.
\eeq 
Thus, applying the $S$-correspondence to a Taylor series we find that at the points obeying the
conditions (\ref{condiciones}) there is a cut-off in the series and, hence, the sums will always
converge in case the function has a well defined parity.

In the following diagram we display the basis elements $(\xi^n, \Delta,x^{(n)})$ of the three usual
representations: 

$$
\begin{array}{c@{}c@{}c@{}c@{}c}
\text{\bf Right representation}& &
\text{\bf Left representation}& &
\text{\bf Symmetric representation}\\[0.4cm]
\:\:\xi_+^n=\left(X T^{-1}\right)^n&\overset{\mathcal{R}}{\longleftrightarrow}&
\:\:\xi_-^n=\left(X T\right)^n&\overset{\mathcal{R}}{\longleftrightarrow}&
\:\:\xi_s^n=\left(2X \left(T+T^{-1}\right)^{-1}\right)^n\\[5pt]
\mathfrak{J}\updownarrow& &\mathfrak{J}\updownarrow& &\mathfrak{J}\updownarrow\\[5pt]
\:\:\dps\Delta_+=\frac{1}{\sigma}\left(T-1\right)&\overset{\mathcal{R}}{\longleftrightarrow}&
\dps\Delta_-=\frac{1}{\sigma}\left(1-T^{-1}\right)&\overset{\mathcal{R}}{\longleftrightarrow}&
\dps\Delta_s=\frac{1}{2\sigma}\left(T-T^{-1}\right)\\[5pt]
\mathfrak{J}\updownarrow& &\mathfrak{J}\updownarrow& &\mathfrak{J}\updownarrow\\[5pt]
x_+^{(n)}
=\dps\prod_{i=0}^{n-1}(x-i\sigma)
&\overset{\mathcal{R}}{\longleftrightarrow}&
x_-^{(n)}
=\dps\prod_{i=0}^{n-1}(x+i\sigma) 
&\overset{\mathcal{R}}{\longleftrightarrow}&
x_s^{(n)}=
\dps x\prod_{i=0}^{n-2}(x+(2i-n+2)\sigma) .
\end{array}
$$

%
%
%
%
%


\begin{figure}[t]

\vskip-1cm

\centering
\subfigure[Some powers for the position in the S-correspondence]
{\includegraphics[width=0.65\textwidth]{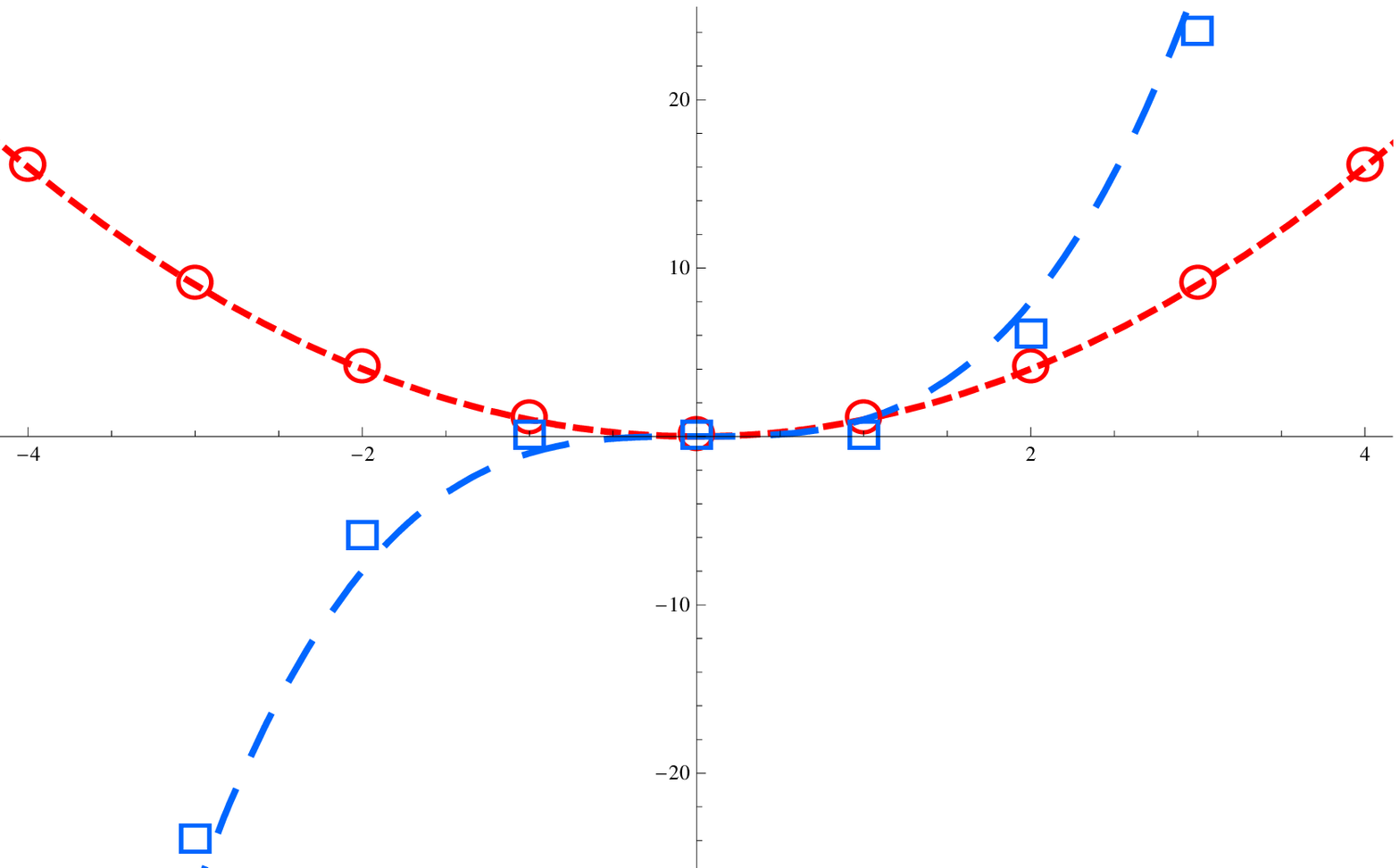}\label{DibPotenciaPosSim}}\\
\subfigure[Some powers for the position in the left-correspondence]
{\includegraphics[width=0.47\textwidth]{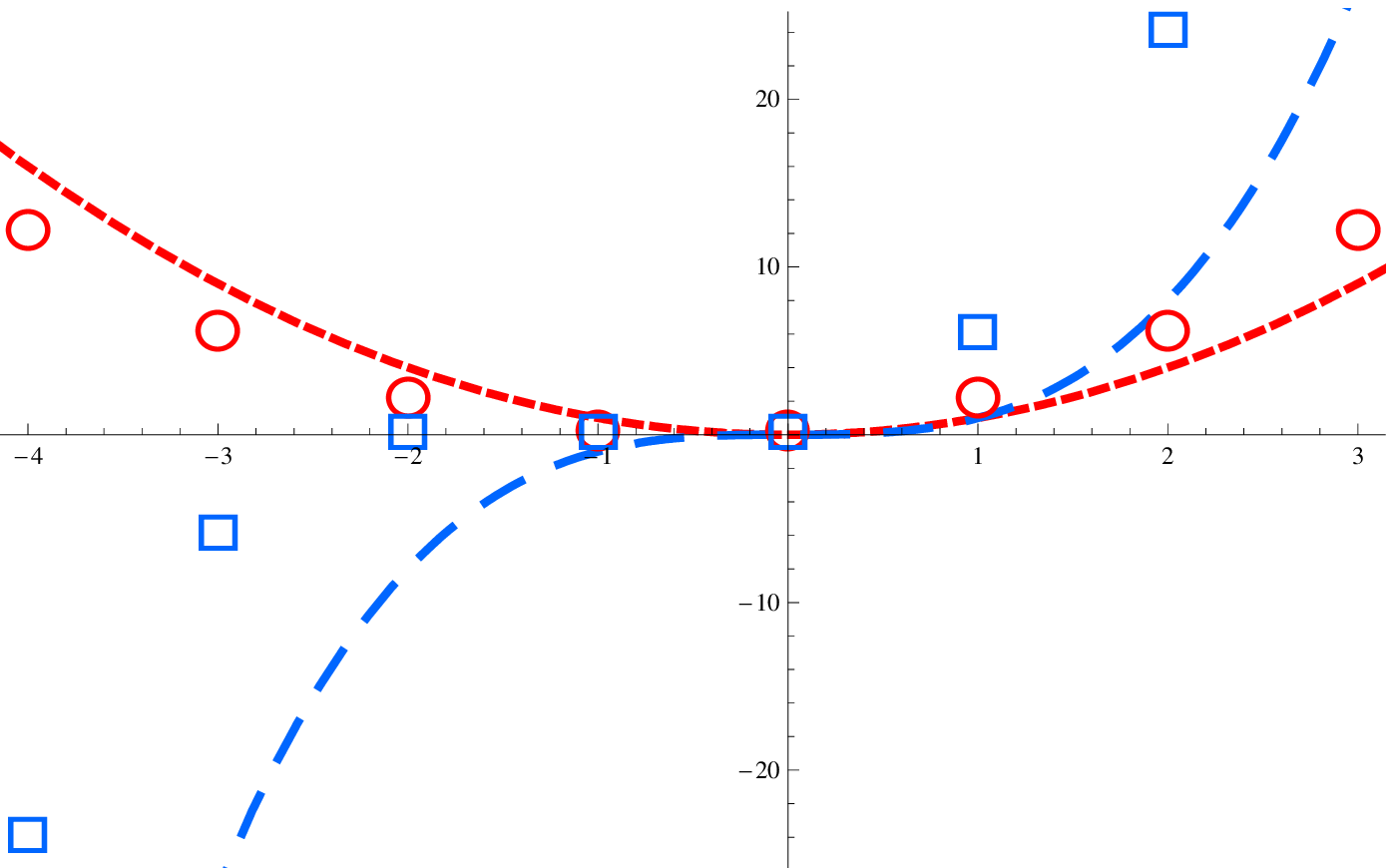}}
\hskip0.02\textwidth
\subfigure[Some powers for the position in the right-correspondence]
{\includegraphics[width=0.47\textwidth]{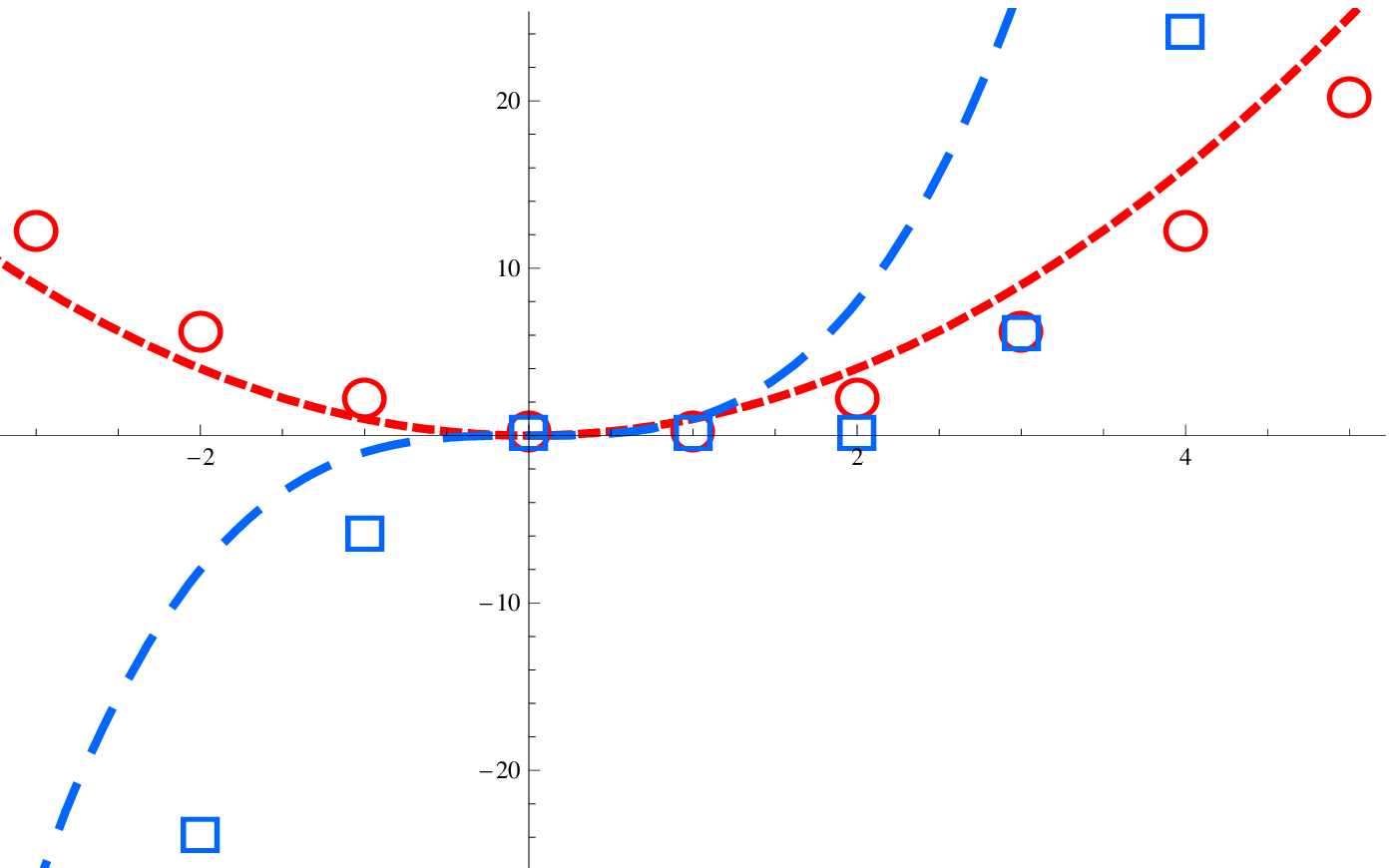}}

\caption{\footnotesize Represented in dashed line the continuous powers and in dots the powers for
the discrete position with red circles for $x^2$ and blue squares for $x^3$. The number of
zeros is equal to the power $n$ for the right and the left cases.}

\label{DibPotenciaPosID}

\end{figure}



\section{The discretized exponential function and the trigonometric
functions}\label{discreteexponential}

\begin{figure}[t]
\vskip-1cm
\centering
\subfigure[Exponential function with $\sigma=0.2$.]{\label{DibExp}\hskip-1cm
\includegraphics[width=0.6\textwidth]{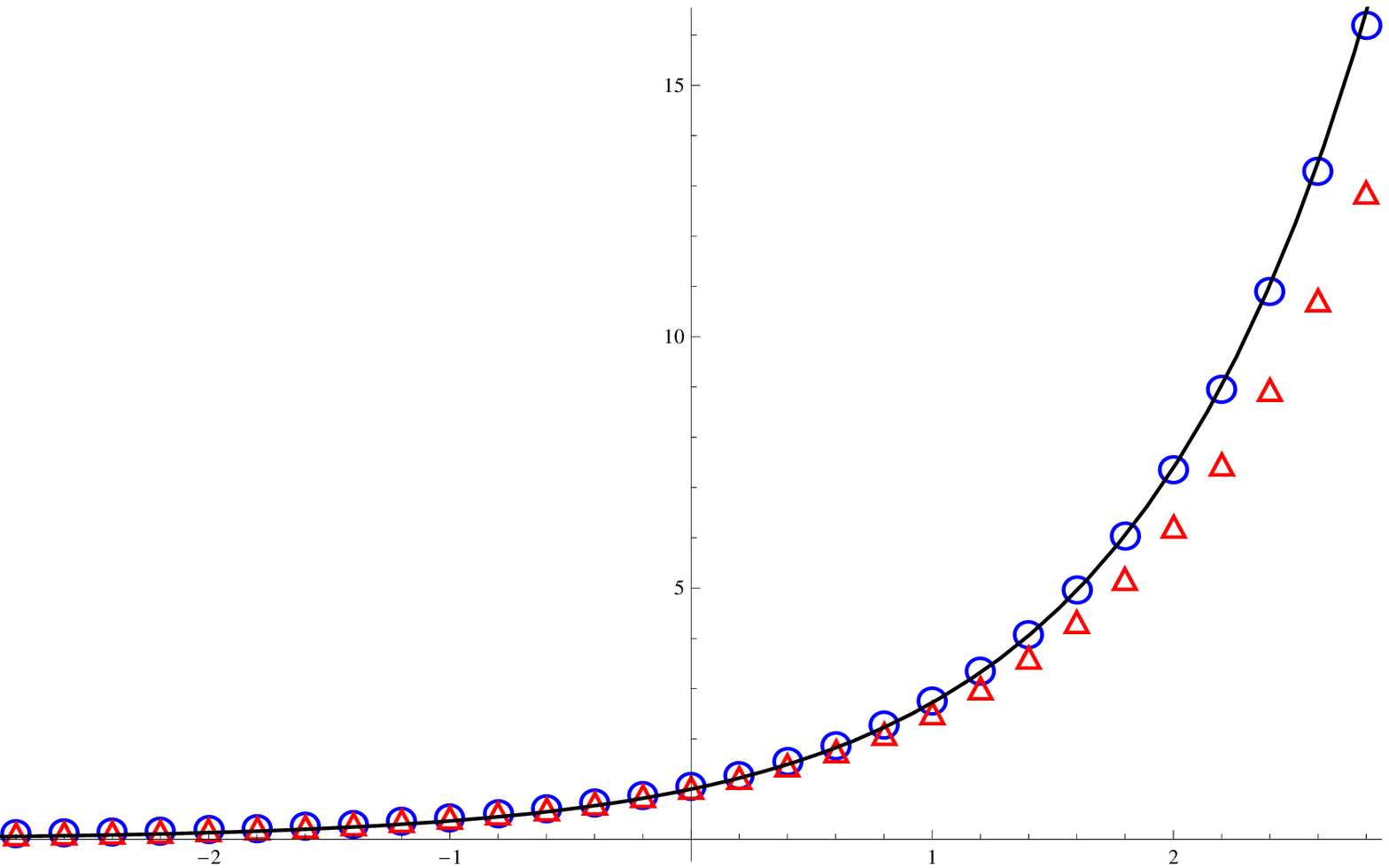}\hskip-7cm
\psframe[fillstyle=solid,fillcolor=white,linecolor=white](-4,5)(4,1.5)\hskip3cm
\includegraphics[width=0.215\textwidth,viewport=400 -130 630 200]
{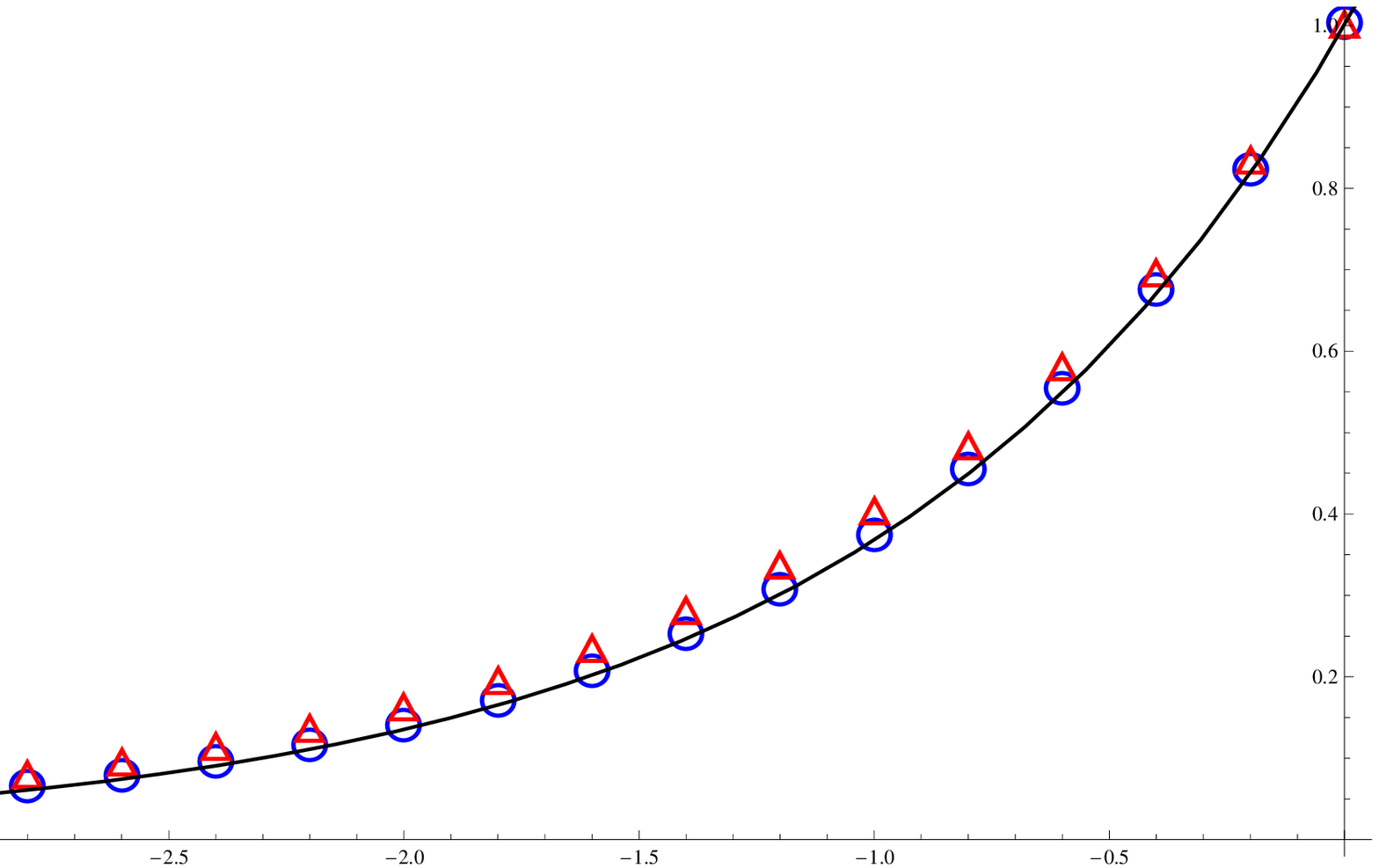}}
\\
\subfigure[Sinus function with $\sigma=0.3$]
{\label{DibSin}\includegraphics[width=0.6\textwidth]{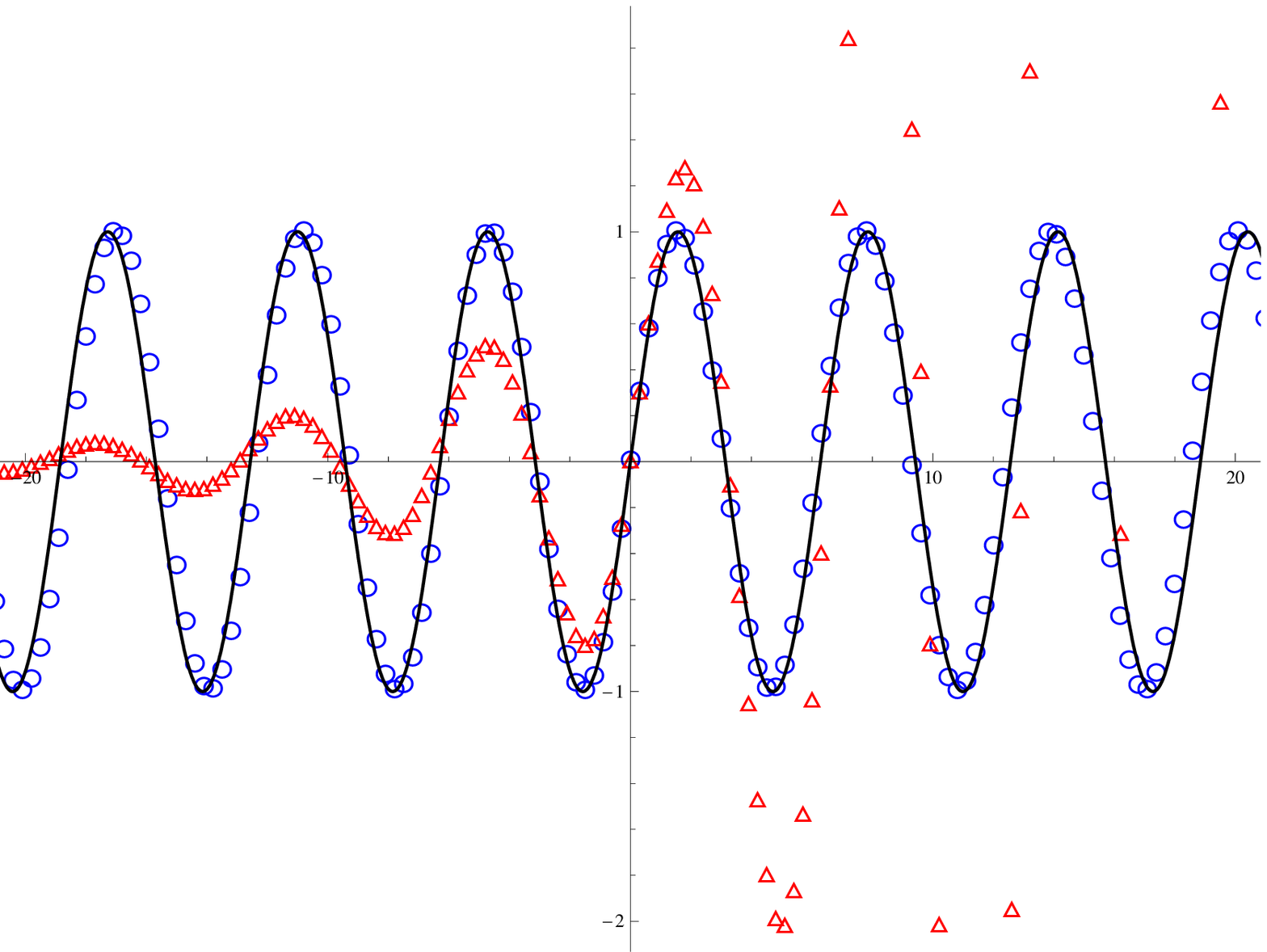}}

\caption{\footnotesize We represent here the continuous functions with a black solid line, and
the discrete ones with red triangles the right correspondences and with blue circles the symmetric
ones.}

\end{figure}

We start with the exponential function which in some sense, is the simplest case.
The continuous exponential function, as it is well known, is defined through the following
differential equation
\[
\frac{\dif}{\dif x}\ y_k=k y_k,\quad y_k(0)=1, \qquad k\in \mathbb{R},
\]
and its power expansion series around the origin is
\[
y_k(x))=e^{k x}=\sum_{n=0}^\infty \frac{(k x)^n}{n!} .
\]
\ind
As we mentioned before, the umbral correspondence gives rise to the same function either by solving
the discrete difference equation, obtained by applying the umbral correspondence to the continuous
differential equation, or by applying the umbral correspondence to  its
solutions as long as their power series converge. Hence, we have to calculate 
\beq
\exp_U(k, x)=\sum_{n=0}^\infty \frac{k^n}{n!}\xi^n \cdot 1\; .
\label{expum}
\eeq
If we use the right-correspondence (\ref{xbetaMas}) the umbral-exponential (\ref{expum}) becomes
\[
\exp_+(k, \sigma m)=\left\{\begin{array}{cl}
\dps\sum_{n=0}^m (k\sigma)^n \frac{m!}{n!(m-n)!} & \text{if }m>0\\
\dps\sum_{n=0}^\infty (-k\sigma)^n
\frac{(-m+n-1)!}{n!(-m-1)!}  & \text{if }m<0
\end{array}\right\}=(1+k\sigma)^m.
\]
For the left-correspondence we get
\[
\exp_-(k, \sigma m)=\left\{\begin{array}{cl}
\dps\sum_{n=0}^\infty (k\sigma)^n
\frac{(m+n-1)!}{n!(m-1)!} & \text{if }m>0\\
\dps\sum_{n=0}^{-m} (-k\sigma)^n \frac{(-m)!}{n!(-m-n)!}  & \text{if }m<0
\end{array}\right\}=(1-k\sigma)^{-m}.
\]
In Fig.~\ref{DibExp} it is displayed the continuous exponential function together with its discrete
version by the right-correspondence. It is worthy to note that if in the continuous case we found
that  $\exp(k\;x)=\exp((-k)(-x))$, now in the discrete case we have that $\exp_+(k, x)=\exp_-(-k,
-x)$.

\ind
Another possibility is to apply the $S$-correspondence (\ref{xbetasn}). So, we obtain
\[
\exp_{s}(k, m\sigma)=\sum_{n=0}^\infty
\frac{(k\sigma)^n}{n!}m\prod_{i=0}^{n-2}
(m+2i-(n-2))=\left(k\sigma+\sqrt{(k\sigma)^2+1}\right)^{m}.
\]
\ind These discrete exponentials only converge when the continuous variable $k$ obeys,
$(k\sigma)^2<1$, for all the three correspondences studied in the previous sections.

\ind
Also, we can easily study the trigonometric functions (as well as the hyperbolic functions) taking
into account their expressions in terms of the exponential functions. In this way, we can obtain
the discrete trigonometric functions from the discrete exponential function. We display in
Fig.~\ref{DibSin} the sine function.

We can see that the symmetric discrete trigonometric functions maintain  the periodic
behaviour of the continuous ones. However the period $\lambda_s$ is different to the continuous
function $\lambda=2\pi/k$. We find it easier in the case of the sine function through
$\sin_s(k,l\sigma)=\sin_s(k,0 \sigma)$. We have fixed $\lambda=l\sigma$ with $l\in\mathbb{Z}$, but
it can be generalized without problems to $l\in\mathbb{R}$.
\begin{gather}
\sin_s(k,l\sigma)=\frac{1}{2\im}\left(\left(\sqrt{1-(k\sigma)^2}+\im k\sigma\right)^l-
\left(\sqrt{1-(k\sigma)^2}-\im k\sigma\right)^l\right)=0=\sin_s(k,0\sigma)\non\\
\Rightarrow\left(\sqrt{1-(k\sigma)^2}+\im k\sigma\right)^l=\left(\sqrt{1-(k\sigma)^2}-\im
k\sigma\right)^l\non
\end{gather}
to solve this identity we define
$$
\sqrt{1-(k\sigma)^2}+\im k\sigma=r e^{\im\theta}\Rightarrow
\left\{ \begin{array}{rl}
\theta=&\arctan\left(\frac{k\sigma}{\sqrt{1-(k\sigma)^2}}\right)\\
         r=&1
\end{array}\right.
$$
and this guide us to the solution
$$
\sin(\theta l)=0 \Rightarrow \theta_n=\frac{n\pi}{l}\quad \Rightarrow\quad
k_n=\frac{1}{\sigma}\sin \frac{n\pi}{l}.
$$
As there are two zeros in each period of the function, we can find the relation between the number
of waves and the wave length. We have added an superscript to remark that these relations are
different for each representation, then
\beq
k^s=\frac{1}{\sigma}\sin\left(\frac{2\pi}{l}\right)\:,\quad\lambda^s=\frac{2\pi\sigma}{
\arcsin(k^s\sigma)}.
\label{EqMomLonOndaSim}
\eeq
It is easy to demonstrate the periodic behaviour of the symmetric trigonometric functions, in fact
introducing (\ref{EqMomLonOndaSim}) into de symmetric sinus
$$
\sin_s(k^s,m\sigma)=\sin\left(m\frac{2\pi}{l}\right).
$$
\ind
For the right and left correspondences the trigonometric functions are not periodic. However the
extremums of the functions and the intersection points with the abscissa axis appear periodically
and then we can define a wave function through this property. With a similar computation than in the
symmetric case, we find
\beq
k^+=\frac{1}{\sigma}\tan\left(\frac{2\pi}{l}\right)\:,\quad\lambda^+=
\frac{2\pi\sigma}{\arctan(k^+\sigma)}.
\label{EqLonOndaMas}
\eeq
\ind
The wave length is bigger in the right and left discretizations and smaller in the symmetric
discretization than in the continuous trigonometric functions. Also in these three
correspondences the trigonometric functions have a minimum period, it appears
in the convergence limit, when $k\sigma=1$. For the right and the left discretizations we have 
$\lambda^+_{min}=8\sigma$, in fact it is a 8 point wave, and for the
symmetric one $\lambda^s_{min}=4\sigma$, in fact it is a 4 point wave.

We set $\lambda=l \sigma$ and without loss of generality we fix $l$ integer. Then it is easy to
calculate the change in the amplitude in the right and left correspondences through
(\ref{EqLonOndaMas})
$$
\sin_+(k^+,m\sigma)=\frac{1}{2\im}\left((1+\im k^+\sigma)^m-(1-\im k^+\sigma)^m\right)=
\cos^{-m}\left(\frac{2\pi}{l}\right)\sin\left(m\frac{2\pi}{l}\right)
$$
and
$$
\sin_+(k, m\sigma)A_n(l)=
\sin_+(k, (m+n l)\sigma)
$$
Hence, the non periodic part is then the multiplicative factor
\beq
A_n(l)=\left(\sec^l\left(\frac{2\pi}{l}\right)\right)^n.
\label{EqCambioAmpli}
\eeq
This is also true for the cosine function and for any $l\in \mathbb{R}$. Although we have a wave
length $\lambda=l \sigma$ for the trigonometric functions in the right and left representations, we
can see that they are not periodic.

It is important to note that all the properties that obey the exponential and the trigonometric
functions are maintained after the discretization with an exception: any property that includes two
or more different constants $k$, because they give similar effects than a different lattice spacing
$\sigma$, nor a different $x$ like in the continuous case. Thus, for example
\bea
&\exp_U(k, m\sigma)\exp_U(k, n\sigma)=\exp_U(k, (m+n)\sigma),\non\\[4pt]
&\exp_U(k, m\sigma)\exp_U(k', m\sigma)\neq\exp_U(k+k', m\sigma)\non.
\eea

%
%
%



\section{Umbral Schr\"odinger equation}

Let us go to discretize the Schr\"odinger equation with a potential $V(x)$. We will profit the
well known fact that if the potential is independent of the time then the continuous Schr\"odinger
equation is separable. Thus, from the Schr\"odinger equation
\[
\im\hbar\partial_t\phi(x,t)=-\frac{\hbar^2}{2m}\partial_x^2 \phi(x,t)+V(x)\phi(x,t) ,
\]
and considering $\phi(x,t)=\psi(x)\varphi(t)$ we obtain two uncoupled ordinary differential equations 
\[\begin{array}{c}
\dps-\frac{\hbar^2}{2m}\partial_x^2 \psi (x)+V(x)\psi (x)=E\psi(x) ,\\[0.4cm]
\im\hbar\partial_t\varphi(t)=E\varphi(t),
\end{array}\]
that we can rewrite with natural units as
\beq
\begin{array}{c}
\partial_x^2 \psi (x)+V(x)\psi(x)=E\psi(x),\\[0.4cm]
-\im\partial_t\varphi(t)=E\varphi(t) .
\end{array}
\label{equatsep2}
\eeq
\ind
The solution of  the equation (\ref{equatsep2}b) is
\beq
\varphi(t)=C e^{\im E  t} .
\label{EqFuncionOndaTemporal}
\eeq
\subsection{Constant potential}
\smallskip
In relation with the first equation of (\ref{equatsep2}), the  simplest case appears when the
potential is constant ($V(x)=V_0$). The  solution  is
\beq
\begin{array}{lll}
\psi(x)= A e^{\im kx}+B e^{-\im kx}\qquad&
\text{if }\;E>V_0 ,\\[0.3cm]
\psi(x)=A e^{kx}+B e^{-kx}   \qquad&
\text{if }\;E<V_0 ,
\label{EqFuncionOndaEspacial}
\end{array}
\eeq
with $k=\sqrt{|E-V_0|}$. Therefore, the solution $\phi(x,t)$ is a linear combination  of
trigonometric or exponential functions and, hence, their discretizations have been studied in
section~\ref{discreteexponential}.

The energy, for a fixed momentum $k$, is the same than in the continuous case for any
discretization as we can see:
\begin{eqnarray*}
\hskip-2.5cm H\psi &=&\left(-\Delta^2+V_0\right)
\sum_{n=0}^{\infty}\frac{(\im k)^n}{n!}\left(A+(-1)^n B\right)\xi^n=\non\\
\hskip-2.5cm           &=&\left(k^2+V_0\right)\sum_{n=0}^{\infty}
\frac{(\im k)^n}{n!}\left(A+(-1)^n B\right)\xi^n=\left(k^2+V_0\right)\psi.
\end{eqnarray*}
In spite of this result, the relation between the momentum and the wave length $\lambda=l\sigma$ is
different in the discrete case:
$$
k=\frac{2\pi}{\lambda}\:,\quad k_{\pm}=\frac{1}{\sigma}\tan\left(\frac{2\pi}{l}\right)
\:,\quad k_s=\frac{1}{\sigma}\sin\left(\frac{2\pi}{l}\right).
$$
Then the phase velocity changes, although the group velocity is the same
as in the continuous case and, for this reason, the relation between the energy and the velocity.

Let be $t=\tilde{n}\tau$ where $\tilde{n}\in\mathbb{Z}$ the discrete time, and $\tau$ the
fundamental time length. Following the limits of convergence of the discrete exponentials we find
two upper limits for the energy, one in relation with the temporal part of the wave function
(\ref{EqFuncionOndaTemporal}) and the other one in relation with the spatial part of the wave
function (\ref{EqFuncionOndaEspacial}).
\beq
E_{max_t}=\frac{\hbar}{\tau}\: , \quad E_{max_l}=\frac{\hbar^2}{2m\sigma^2}.
\eeq
In fact, we have to take the minor of them in each situation. Taking
$\sigma\thickapprox l_{Planck}=1.62\:10^{-35} m$ and $\tau\thickapprox
t_{Planck}=5.39\:10^{-44} s$ we find values of the maximum energy for the time wave $E_{max_t}$ and
for the spatial wave for a proton $E_{p^+,max_l}$ and an electron $E_{e^-,max_l}$
\begin{eqnarray*}
E_{max_t}\thickapprox E_{Planck}=1.22\:10^{28}eV&
\: , \quad& E_{max_l}=1.38\:10^{20}\frac{1}{m} eV\non\\
E_{p^+,max_l}=7.94\:10^{46}eV&
\: , \quad&E_{e^-,max_l}=1.46\:10^{50}eV\non.
\end{eqnarray*}
\newpage
\noindent
Remark that these are non-relativistic effects. We can study also the change in the amplitude for
the right and left correspondences. The factor of expansion or contraction is
$$
A_n(l)=\left(\sec^l\left(\frac{2\pi}{l}\right)\right)^n\simeq\left(1+\frac{2\pi^2}{l}\right)^n
$$
where $l$ is the number of points into the wave length and $n$ is the number of wave lengths along
the wave. For the temporal part of the wave function of a particle which travel
into the LEP, that is during $9\:10^-5 s$ with an energy of $1.99\:10^{14}eV$, we find that the
factor
is $A_n(l)\sim10^{10^5}$. For the spatial part of the same particle, an electron, the
factor is of order $A_n(l)\sim3$. These factors grow a lot with the energy or the mass of the
particle. As at this energy the relativistic effects are important and we are using now non
relativistic quantum mechanics, it is comprehensible to find these too big numbers.

\subsection{Infinite potential well}
\smallskip
A little more complicated is the infinite potential well, although it is very similar to a constant
potential, it reduces the space to a finite region and due to this fact new differences will appear
between the continuous and the discrete case. The potential is
$$
V(x)=\left\{\begin{array}{cl}
0      & \text{if }\ x\in[0,L]\\[3pt]
\infty & \text{if }\ x\notin[0,L]
\end{array}\right.
$$
\ind
The wave functions must obey the boundary conditions $\psi(0)=\psi(L)=0$ which lead to a countable
number of solutions:
$$
\psi_n(x)=\sin \left(n\pi\frac{x}{L}\right)\: ,\quad
k=\frac{n\pi}{L} \:,\quad n\in\mathbb{Z},
$$
in the same way we apply the boundary conditions to all our discretizations. We can find the quantum
rule for both the right and left correspondences following the computation of
(\ref{EqMomLonOndaSim}). So
$$
k^+_n=\frac{1}{\sigma}\tan \pi\frac{n}{M}
$$
where $L=M\sigma$. Therefore there are only $M/2$ possible states instead of the infinite states of
the continuous case, with a degeneration relative to the parity. The energy is $E_n=k_n^2$ and it is
bounded at $n=(M-1)/2$. So the maximum energy is
$$
E^+_{max}=\frac{\hbar^2}{2m\sigma^2}\tan^2 \left(\pi\left[\frac{M-1}{2M}\right]\right)
$$
At $n=M/2$ we find a non physical state of infinite energy and infinite wave function amplitude.
We can do a small computation finding that an electron in an infinite well of wide the Bohr radius
($M\sim10^{27}$) will have a maximum state energy of $E^+_{max}\sim10^{77}eV$.  The
maximum energy grows when increasing the number of points  $M$ and also when decreasing the mass of
the particle. But the wave function, as we saw previously, has a convergence range that limits
the number of states to $n\in(-M/4,M/4)$ because $k<1/\sigma$. Hence, the maximum energy
is much more smaller.

\begin{figure}[b]

\vskip-8cm

\centering

\begin{tabular}{cc}
\hskip-0.25cm
\multirow{3}{*}[3.2cm]{\includegraphics[width=0.15\textwidth,
height=0.33\textheight]{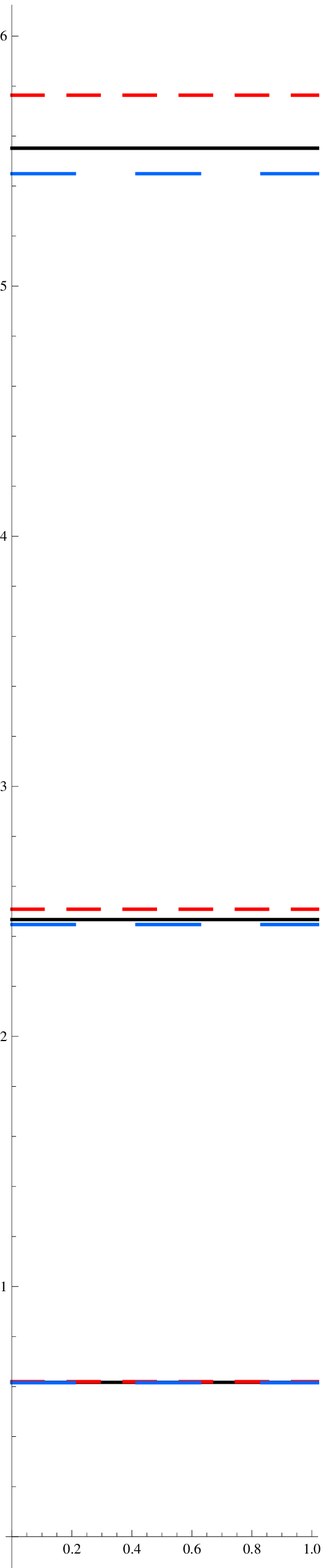}}
&\includegraphics[width=0.82\textwidth]{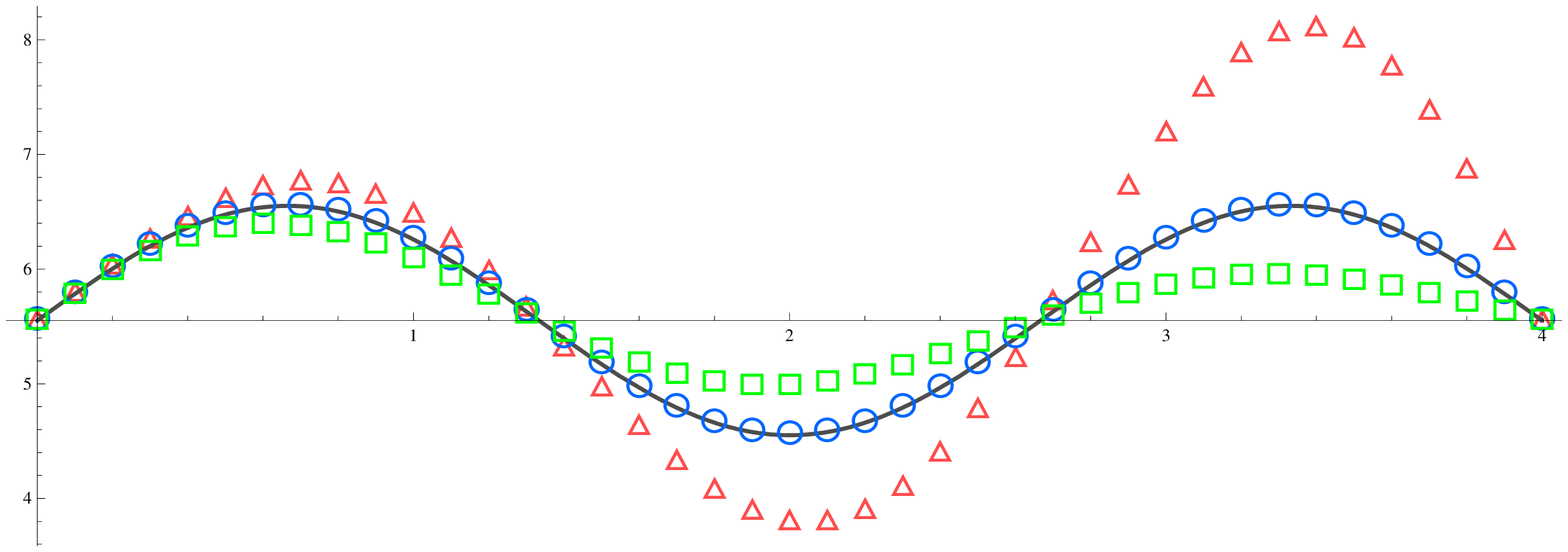}\\
&\includegraphics[width=0.82\textwidth]{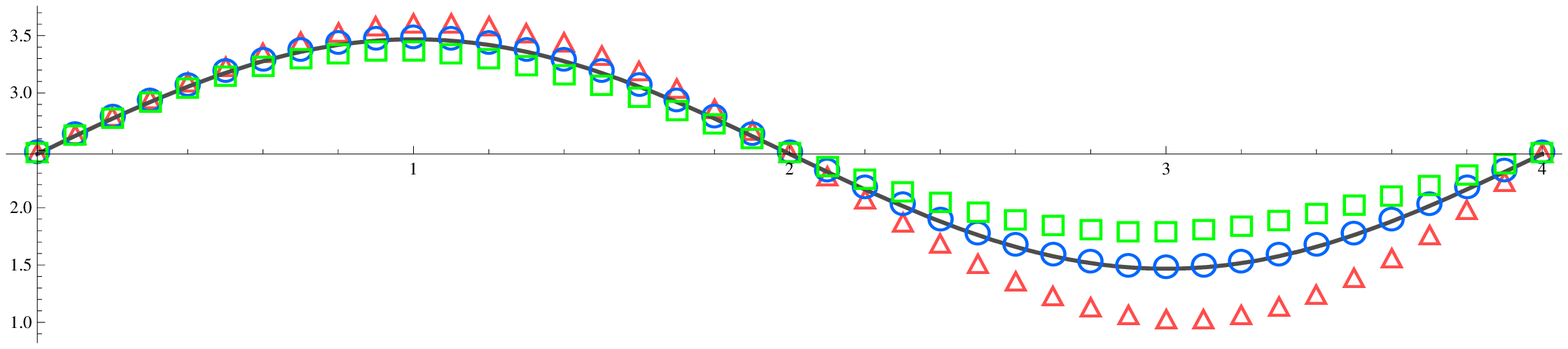}\\
&\includegraphics[width=0.82\textwidth]{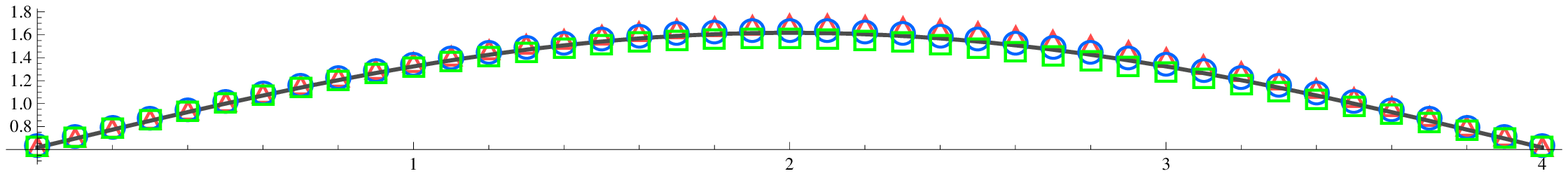}
\end{tabular}

\caption{Energy levels and wave functions of the infinite well. The energy levels are in the left,
ones in solid line for the continuous case and dashed for the discrete case, the shorter red lines
are the right/left case energies and the longer blue lines are the symmetric case energies.
Wave functions are in the right, in black solid line are for the continuous case, blue
circled points are for the S-correspondence, red triangled ones for the right correspondence and
green squared ones the left correspondence.}

\label{pozoinfinito123}

\end{figure}

We have that the quantum rule for the $S$-correspondence is
$$
k^s_n=\frac{1}{\sigma}\sin \pi\frac{n}{M}.
$$
Again we have a finite number of states, $M/2$, rounding down, with a degeneration relative to the
parity also. The energy is always bounded by $\hbar/(2m\sigma^2)$, just as we found with the
convergence condition
$$
E^s_n=\left(k_n^s\right)^2=\frac{1}{\sigma^2}\sin^2 \pi\frac{n}{M}.
$$
\ind
In Figure \ref{pozoinfinito123} are drawn the energy levels and the wave functions of the ground
state and the two first excited states for the continuous case and the three discretizations. We
can see the great fit of the symmetric one for the first excited states, but remarking that as its
number of states is finite, $M/2$, then upper these wave functions there are not more states.
The right and the left ones have an asymmetry to the left and right directions of the space and then
are much more different to the continuous wave functions.
\vskip-3cm


\section{Conclusions}

In this paper we have shown how to apply the techniques of the Umbral Calculus for the
discretization of equations preserving the point symmetries. We can see the advantage of a tool that
may be used not only on the equations, but also over  their solutions. Due to this fact we have
not to solve difference equations because we alternatively can sum series. Moreover,
in the most difficult cases we can use numerical methods to get a solution.

In the umbral discretization we can choose between infinite umbral representations. However, still
in the simplest cases they present very  different behaviours. We can separate all of them in two
classes, one preserving the parity of the equations and solutions (like the $S$-representation) and
the other one that does not (like the right and  left representations). From the point of view of
the local symmetries in \mbox{Ref~(\cite{dimakis}--\cite{salgado}, \cite{levi3})} it was proved that
independently of the correspondence we get the same discrete point  symmetries.

The study of the discrete Schr\"odinger equation together with other potentials (Coulomb,
harmonic oscillator, \dots) are in progress \mbox{\cite{lopez-sendino1}--\cite{lopez-sendino2}} and
the results will be published elsewhere.


\section*{Acknowledgments}
This work was partially supported  by the Ministerio de Educaci\'on y Ciencia  of Spain (Projects
FIS2005-03989 and  by the Junta de Castilla y Le\'on   (Project VA013C05). JLS thanks the
Universidad de Valladolid and the Banco Santander Central Hispano by the  PhD grant of the program
``Ayudas de la Universidad de Valladolid cofinanciadas por el BSCH''.




\section*{References}

\begin{thebibliography}{99}

\bibitem{gibbs} Gibbs P 1995 {\it The small scale structure of space-time: a bibliographical
review} hep-th/9506171

\bibitem{dimakis} Dimakis A, M\"uler-Hoissen F and Striker T 1996 Umbral calculus,
discretization, and quantum mechanics on a lattice {\it J. Phys.} A {\bf 29} 6861

\bibitem{levi1} Levi D, Vinet L and Winternitz P 1997 Lie group formalism for difference
equations {\it J. Phys.} A {\bf 30} 633

\bibitem{olmo1} Levi D, Negro J and del Olmo M A 2001 Discrete derivatives and symmetries of
difference equations {\it J. Phys.} A {\bf 34} 2023

\bibitem{olmo2} Levi D, Negro J and del Olmo M A 2001 Lie symmetries of difference
equations {\it Czech. J. Phys.} {\bf 51} 341

\bibitem{senosiain} Senosiain M J 2002 {\it Teor\'ia de las Ecuaciones Gen\'ericas: Una
Introducci\'on al C\'alculo Umbral} Memoria de Grado Universidad de Salamanca

\bibitem{olmo3} Levi D, Negro J and del Olmo M A 2004 Discrete q-derivatives and symmetries
of q-difference equations {\it J. Phys.} A {\bf 37} 3459--3473

\bibitem{levi2} Levi D, Tempesta P and Winternitz P 2004 Umbral calculus, difference
equations and the discrete Schr\"odinger equation {\it J. Math. Phys.} {\bf 45} 4077

\bibitem{salgado} Salgado E 2004 {\it Mec\'anica Cu\'antica Discreta} Trabajo Fin de Carrera de
F\'isicas Univ. de Valladolid

\bibitem{mullin} Mullin R and Rota G C 1970 {\it On the foundations of combinatorial theory III.
Theory of binomial enumeration} ({\it Graph theory and its applications}) Ed. B. Harris
(Academic Press) p~167-213

\bibitem{blissard1} Blissard J 1861 Theory of generic equations {\it Quart. J. Pure Appl. Math.}
{\bf 4} 279

\bibitem{blissard2} Blissard J 1862 Theory of generic equations {\it Quart. J. Pure Appl.
Math.}
{\bf 5} 58, 184

\bibitem{bell} Bell E T 1938 The history of Blissard's symbolic calculus, with a sketch of
the inventor's life {\it Amer. Math. Monthly} {\bf 45} 414

\bibitem{appell} Appell P 1880 Sur une classe de polyn\^omes {\it Ann. Sci. Ecole Norm. Sup.
(2)} {\bf 9} 119

\bibitem{pincherle} Pincherle S and Amaldi S 1901 {\it Le Operazioni Distributive e le loro
Applicazioni all'Analisi} (Bologna: N. Zanichelli)

\bibitem{davis} Davis H T,  1936, {\it The Theory of Linear Operators} (Bloomington: Principia
Press)

\bibitem{sheffer} Sheffer I M 1939 Some properties of polynomial sets of type zero {\it Duke
Math. J.} {\bf 5} 590

\bibitem{rota1} Rota G C, Kahaner D and Odlyzko A 1973 On the foundations of combinatorial
theory VII. Finite operator calculus {\it J. Math. Anal. Appl} {\bf 42} 684

\bibitem{rota2} Rota G C 1975 {\it Finite Operator Calculus} (San Diego: Academic)


\bibitem{roman2} Roman S M 1982 The theory of umbral calculus I {\it J. Math. Anal. Appl.}
{\bf 87} 58

\bibitem{roman3} Roman S M 1982 The theory of umbral calculus II {\it J. Math. Anal. Appl.}
{\bf 89} 290

\bibitem{roman4} Roman S M 1983 The theory of umbral calculus III {\it J. Math. Anal. Appl.}
{\bf 95} 528

\bibitem{roman5} Roman S M 1975 {\it The Umbral Calculus} (San Diego: Academic)

\bibitem{bucchianico} Di Bucchianico A and Loeb D 2000 A selected survey of umbral calculus
{\it Electr. J. Combin.} DS3

\bibitem{zachos} Zachos C K 2007 Umbral deformations on discrete spacetime {\it Preprint}
hep-th/07102306

\bibitem{levi3} Levi D and Winternitz P 2005 Continuous symmetries of difference equations {\it J.
Phys.} A {\bf 39} R1--R63

\bibitem{levi4} Levi D and Petrera M 2007 Continuous symmetries of the lattice potential KdV
equation {\it J. Phys.} A {\bf 40} 4141--4159

\bibitem{levi5} Levi D, Tremblay S and Winternitz P 2007 Lie point symmetries of difference
equations and lattices {\it Preprint} math-ph/07093112

\bibitem{levi6} Levi D, Tremblay S and Winternitz P 2007 Lie symmetries of
multidimensional difference equations {\it Preprint} math-ph/07093238

\bibitem{levi7} Hernandez Heredero R, Levi D, Petrera M and Scimiterna C 2007 Multiscale expansion
on the lattice and integrability of partial difference equations {\it Preprint} math-ph/07105299

\bibitem{lopez-sendino1} L\'opez-Sendino J E, Negro J, del Olmo M A and Salgado E 2008
Discretizing quantum mechanics using umbral calculus {\it Preprint}

\bibitem{lopez-sendino2} L\'opez-Sendino J E, Negro J and del Olmo M A 2008 Discrete harmonic
oscillator and umbral calculus {\it Preprint}

\end{thebibliography}
\end{document}